\documentclass{article}

\usepackage{arxiv}

\usepackage[utf8]{inputenc}
\usepackage[T1]{fontenc}
\usepackage{url}
\usepackage{booktabs}
\usepackage{amsmath}
\usepackage{amsfonts}
\usepackage{amssymb}
\usepackage{nicefrac}
\usepackage{microtype}
\usepackage{graphicx}
\usepackage{multirow}
\usepackage{xcolor}
\usepackage{natbib}
\usepackage{hyperref}

\title{A Classroom Study of LLM-generated Feedback Intervention in Introductory Programming}

\author{%
  Hasnain Heickal \\
  Manning College of Information and Computer Sciences \\
  University of Massachusetts Amherst \\
  \texttt{hheickal@cs.umass.edu} \\
  \And
  Andrew Lan \\
  Manning College of Information and Computer Sciences \\
  University of Massachusetts Amherst \\
  \texttt{andrewlan@cs.umass.edu} \\
}

\renewcommand{\shorttitle}{LLM Feedback in Intro Programming}

\hypersetup{
  pdftitle={A Classroom Study of LLM-generated Feedback Intervention in Introductory Programming},
  pdfauthor={Hasnain Heickal, Andrew Lan},
}

\begin{document}

\maketitle

\begin{abstract}
Large language models (LLMs) are increasingly used to provide automated feedback in introductory programming courses, yet empirical evidence from authentic classroom deployments comparing different feedback modalities remains limited. In this work, we present a large-scale classroom study in which AI-generated feedback was deployed through a randomized protocol in an introductory Python programming course. Students received one of three feedback conditions on incorrect submissions: natural language hints, AI-generated failing test cases, or no AI feedback. We release the resulting dataset, \textsc{ProgFeed},\footnote{Publicly available at \url{https://github.com/umass-ml4ed/progFeed-dataset-public}.} which captures 6{,}693 submissions from 215 consenting students across 17 labs, including feedback conditions, execution-based performance measures, and fine-grained temporal information. Using this data, we analyze learning trajectories, feedback quality, and submission behavior over repeated attempts. We find that natural language feedback is significantly associated with higher completion rates and faster convergence to correct solutions. Test case feedback, by contrast, exhibits heterogeneous effects that depend critically on feedback validity. Our results suggest that the form of AI-generated feedback matters, and that evaluating feedback quality---not just its presence---is essential for understanding its pedagogical impact.
\end{abstract}


\section{Introduction}

Feedback plays a central role in human learning, enabling learners to identify mistakes, refine strategies, and make progress toward mastery. In classroom settings, however, providing timely and individualized feedback is challenging: instructors must balance limited time and attention across large and heterogeneous student populations. Prior work has shown that immediate, just-in-time feedback \cite{Johnston2015TheEO} and personalized feedback \cite{Kochmar2020AutomatedPF} can significantly improve learning outcomes, yet such feedback is difficult to deliver consistently at scale. This challenge is particularly salient in introductory programming courses, where students require repeated guidance while debugging syntax and logic errors.

Automated feedback systems offer a promising avenue for scaling instructional support. Due to the structured nature of programming languages, a substantial body of work has explored automated feedback for programming tasks, including syntax error explanations, bug localization, and hint generation \cite{keuning2018systematic}. More recently, advances in large language models (LLMs) have expanded the range of feedback that can be generated automatically, from fluent natural language explanations to dynamically constructed test cases \cite{kazemitabaar2023studying, kumar2024testcase}. As LLM-based feedback becomes increasingly integrated into real classrooms, a key open question is how different forms of AI-generated feedback affect student performance and learning over time.

Addressing this question requires data that captures not only student code but also the feedback received and how students respond across multiple attempts. Large-scale programming datasets such as CodeNet \cite{puri2021codenet} lack controlled feedback variation, making it difficult to disentangle feedback effects from confounding factors such as student ability or problem difficulty. Longitudinal datasets like Blackbox \cite{brown2014blackbox} capture rich interaction traces, but without experimental manipulation. In this work, we address this gap directly.

We report on a classroom study in an introductory Python programming course in which AI-generated feedback was deployed through a randomized protocol and delivered \textit{in situ} conditioned on each submission. The resulting dataset, \textsc{ProgFeed}, links complete submission histories to feedback conditions, execution outcomes, and temporal ordering, supporting analyses of short-term performance, time-to-success, and iteration behavior. Our contributions are:
\begin{itemize}
    \item A classroom study enabling controlled comparison of natural language, failing test case, and no-AI feedback conditions in an authentic introductory programming course.
    \item \textsc{ProgFeed}, a longitudinal dataset capturing fine-grained submission trajectories, feedback conditions, and execution outcomes from this study.
    \item Empirical analyses demonstrating how the dataset supports rigorous study of learning dynamics and feedback effectiveness.
\end{itemize}

\section{Related Work}

\subsection{Automated Feedback in CS Education}

Automated feedback systems for introductory programming have traditionally focused on syntax errors and structural issues, leveraging compiler messages, static analysis, or unit tests \cite{keuning2018systematic}. Early approaches provided next-step hints in the edit-distance space \cite{paassen2017continuous} or learned mappings between program representations to generate feedback at scale \cite{piech2015learning}. With the rise of transformer-based models, research has increasingly focused on program repair and error explanations \cite{yasunaga2020graph, yasunaga2021break, leinonen2023syntax}. A key distinction is between \emph{syntax errors}---studied extensively, with systems such as PyFiXV \cite{phung2023syntax} demonstrating strong performance---and \emph{logical errors}, which require understanding program semantics and are far less addressed by automated systems.

\subsection{Logical Errors and LLM-Based Feedback}

Recent work has begun exploring automated feedback for logical errors, particularly using LLMs. One common approach is failing test case feedback, which provides students with counterexamples to guide debugging \cite{kumar2024testcase}. Other approaches generate Socratic questions \cite{kumar2024socratic} or concise natural language hints \cite{phung2023benchmark, phung2023hintgen}. Several studies have evaluated LLM-generated feedback through expert annotation or small-scale user studies \cite{phung2023benchmark, pankiewicz2023feedback}, providing quality insights but typically examining a single modality, few problems, or limited participants---and rarely studying how feedback affects behavior across multiple attempts in authentic settings.

\subsection{Evaluating Feedback Effectiveness}

Evaluating the pedagogical impact of feedback remains challenging. Many studies rely on binary success metrics, though the \emph{process} of learning---iteration patterns, time to resolution, error recurrence---is equally important \cite{ahadi2018learning}. Datasets such as CodeNet \cite{puri2021codenet} provide scale but lack instructional context, while longitudinal datasets like Blackbox \cite{brown2014blackbox} capture traces without controlled interventions. \textsc{ProgFeed} bridges this gap by embedding randomized feedback interventions directly into data collection, linking complete submission trajectories to specific feedback conditions.

\section{Dataset and Study Design}
\label{sec:dataset}

\subsection{Course Context and Participants}
\label{sec:participants}

We conducted an IRB-approved classroom study in an introductory Python programming course at a U.S.\ R1 institution during the Fall 2025 semester. Participation was limited to students aged 18 or older. Of the 365 students enrolled in the course, 220 consented to participate; the 215 consenters who submitted at least one solution constitute the released \textsc{ProgFeed} dataset. Students who did not consent received identical instructional experiences, including access to AI-generated feedback, but their data were excluded from all analyses and from the released dataset.
The course primarily served first-year students, with additional enrollment from second- and third-year students. Most participants were majors in computer science, with representation from related disciplines such as mathematics, physics, and linguistics. The course met twice weekly for 75-minute lectures and included weekly laboratory assignments consisting of auto-graded programming problems.

AI-generated feedback was provided \emph{only} for laboratory assignments. Quizzes, exams, pre-labs, and larger projects did not include AI feedback and are not part of the dataset. In addition to labs, students completed optional midweek pre-lab assignments that were structurally identical to labs and offered as extra-credit practice. Each pre-lab served as both a post-test for the previous lab and a pre-test for the upcoming lab, enabling within-course evaluation of learning across instructional units.

\subsection{Data Collection and Experimental Design}
\label{sec:data_collection}

\textsc{ProgFeed} contains all student submissions to lab and pre-lab problems. Students were allowed unlimited submissions per problem; each was evaluated against fixed instructor-authored test cases, always returning pass/fail verdicts. Feedback assignment was randomized at the submission level within each lab; students were reassigned between labs to prevent systematic advantage. Feedback was \emph{conditionally triggered} only on failing submissions, so exposure is not independent of correctness---all analyses are associative, not causal.

\subsection{Collected Signals}
\label{sec:signals}

\textsc{ProgFeed} is organized hierarchically across students, labs, problems, and submissions. For each submission, the dataset includes submitted source code, submission order within a student--problem sequence, fixed test case outcomes, execution-based performance metrics, and the assigned feedback condition and content when applicable. The dataset also includes auxiliary signals: a pre-course demographic survey, weekly post-lab surveys on perceived feedback usefulness, an end-of-course exit survey, and aggregate course performance measures such as lab scores, quiz scores, and final grades.

\subsection{Privacy and De-identification}
\label{sec:privacy}

All personally identifiable information was removed prior to release. Student identifiers were replaced with deterministic pseudonymous identifiers, preserving longitudinal structure while preventing re-identification. Automated scrubbing was applied to all code and text fields to remove residual identifying information such as names, email addresses, and file paths, while preserving program semantics.

\subsection{Feedback Conditions and Generation}
\label{sec:feedback_conditions}

Each eligible (failing) submission was assigned to one of three conditions:
\begin{description}
    \item[Group 0: No AI Feedback.] Fixed test case pass/fail verdicts only.
    \item[Group 1: Failing Test Case Feedback.] An AI-generated input--output pair designed to fail the student's submission.
    \item[Group 2: Natural Language Feedback.] A concise natural language hint identifying a likely issue, without code suggestions or test cases.
\end{description}

All feedback was generated by \texttt{gpt-4o-mini} via the OpenAI API, given the problem description and student code. A smaller model was chosen for cost reasons given class size. Prompts were fixed across all labs; variation arose solely from student code.

\paragraph{Prompt for Failing Test Case Feedback (Group 1)}
\begin{quote}
\small\texttt{The student is given the problem and asked to implement the specified function(s). You can see their code below. Can you give a test case that would cause the student's code to fail? Only give the test case with the expected output. Do not explain why it fails. Do not include any markdown formatting. If the test case involves a list, use explicit values rather than variable names.\\ Problem: ... Code: ...}
\end{quote}

\paragraph{Prompt for Natural Language Feedback (Group 2)}
\begin{quote}
\small\texttt{The student is given the problem and asked to implement the specified function(s). You can see their code below. Can you give a brief natural language hint explaining why the code is failing? Refer to relevant parts of the problem description if appropriate. Do not suggest code changes. Do not give test cases. Focus on one important issue.\\ Problem: ... Code: ...}
\end{quote}

\subsection{Dataset Statistics}
\label{sec:dataset_stats}

\textsc{ProgFeed} contains 6{,}693 submissions from 215 consenting students across 17 labs and 43 problems. Group 0 (no feedback): 4{,}067 submissions; Group 1 (test-case): 1{,}605; Group 2 (natural-language): 1{,}021. Students submitted multiple attempts per problem, supporting hierarchical analysis at student, problem, and lab levels.

\section{Analysis}

\subsection{Post-Feedback Learning Dynamics}

\begin{figure}[htbp]
    \centering
    \includegraphics[width=0.85\linewidth]{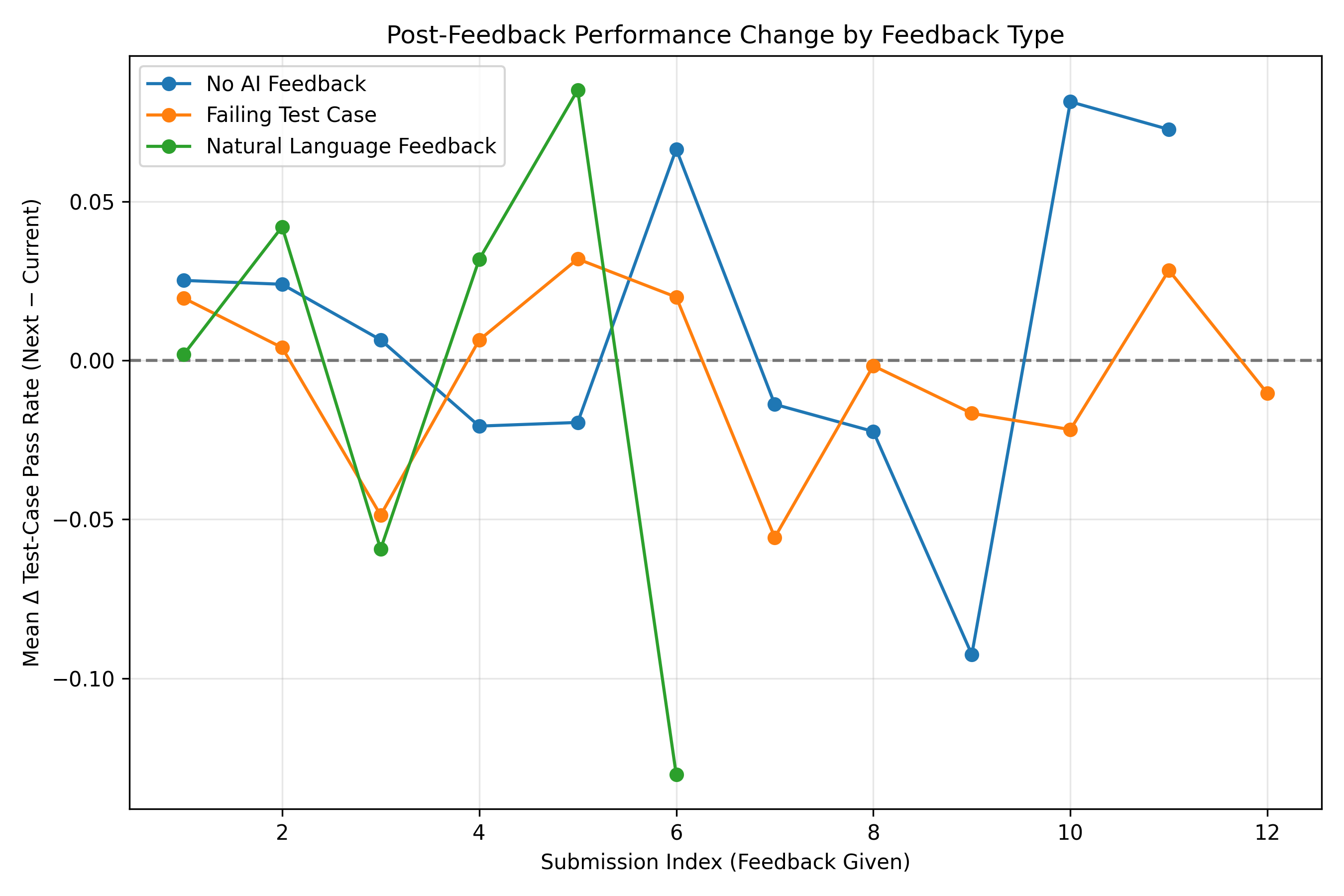}
    \caption{Mean change in test case pass rate by submission index, stratified by feedback condition.}
    \label{fig:trajectories}
\end{figure}

We examine short-term learning dynamics by analyzing $\Delta$ pass rate---the change in test case pass rate between consecutive submissions. Figure~\ref{fig:trajectories} plots mean $\Delta$ pass rate as a function of submission index, stratified by feedback condition. We find that across all conditions, mean $\Delta$ pass rates fluctuate near zero, indicating that immediate improvement following feedback is neither consistent nor guaranteed. Positive gains at some indices are often followed by stagnation or regression at later attempts.

We observe modest early gains under natural language feedback, while test case feedback more frequently exhibits near-zero or negative changes. However, these patterns are not stable: trajectories overlap substantially, and no condition shows uniformly superior short-term improvement. Variability increases at later submission indices, which correspond to a shrinking subset of students who continue submitting after repeated failures---suggesting that feedback effects diminish and students may be attempting randomly at that stage.

\subsection{Baseline Post-Feedback Performance}

We summarize post-feedback performance using $\Delta$ pass rate, computed over valid submission transitions and excluding final submissions with no subsequent attempt. Table~\ref{tab:baseline_delta_performance} reports descriptive statistics by condition. We find that mean $\Delta$ pass rates are near zero across all groups, and fewer than 35\% of transitions result in positive gains. The large standard deviation reflects substantial heterogeneity in short-term responses.

Differences across conditions are modest: natural language feedback yields the highest probability of improvement, while test case feedback shows a slightly higher mean $\Delta$ and more submissions per student--problem on average. We fit a clustered logistic regression predicting improvement ($\Delta > 0$) with standard errors clustered at the student level, and find no significant differences across conditions (Wald test, $p = 0.50$). These patterns motivate our subsequent time-to-success analyses rather than serving as evidence of differential effectiveness at baseline.

\begin{table}[htbp]
\centering
\small
\scalebox{.9}{
\begin{tabular}{lcccc}
\toprule
\textbf{Group} &
\textbf{Mean $\Delta$} &
\textbf{Std.\ Dev.} &
$\boldsymbol{\Pr(\Delta > 0)}$ &
\textbf{Avg.\ Sub} \\
\midrule
No AI      & 0.020 & 0.459 & 0.314 & 1.75 \\
Test Case  & 0.027 & 0.453 & 0.327 & 2.22 \\
Natural Lang & 0.017 & 0.449 & 0.345 & 1.66 \\
\bottomrule
\end{tabular}}
\caption{Baseline post-feedback performance by condition.}
\label{tab:baseline_delta_performance}
\end{table}

\subsection{Effects of AI Feedback}

\begin{table}[htbp]
\centering
\caption{Associations between AI feedback and success and time to correctness}
\label{tab:feedback_effects}
\begin{tabular}{lcccc}
\toprule
\textbf{Comparison} &
\multicolumn{2}{c}{\textbf{Prob.\ of Success}} &
\multicolumn{2}{c}{\textbf{Time to Correctness}} \\
\cmidrule(lr){2-3} \cmidrule(lr){4-5}
 & Coef.\ (SE) & $p$ & HR (95\% CI) & $p$ \\
\midrule
NL vs.\ No AI & $0.84\ (0.30)$ & $0.004$ & $1.09\ [1.01,\ 1.19]$ & $0.03$ \\
TC vs.\ No AI & $0.20\ (0.24)$ & $0.40$  & $0.92\ [0.85,\ 1.00]$ & $0.05$ \\
\bottomrule
\end{tabular}
\end{table}

We investigate whether AI-generated feedback is associated with students' ability to reach a correct solution and the effort required to do so. All analyses are at the student--problem level, including only initially incorrect submissions. We acknowledge that because feedback delivery is conditionally triggered rather than fully randomized, all results are associative rather than causal.

\paragraph{Probability of Eventual Success}
For each student--problem pair, we define a binary outcome indicating whether the student ever reached a fully correct submission. We fit separate logistic regression models comparing each condition to the no-AI baseline with standard errors clustered at the student level. As shown in Table~\ref{tab:feedback_effects}, we find that natural language feedback is significantly associated with higher eventual success ($\beta = 0.84$, $p = 0.004$), corresponding to an odds ratio of approximately 2.33: students receiving natural language feedback are more than twice as likely to eventually reach correctness. We find no significant association for test case feedback ($\beta = 0.20$, $p = 0.40$).

\paragraph{Time to Success}
We model time to correctness using Cox proportional hazards models with right censoring, where time is measured in submissions since first feedback exposure. We find that natural language feedback is associated with a modest but significant increase in the hazard of reaching correctness (HR $= 1.09$, $p = 0.03$): at any given attempt, students receiving natural language feedback are approximately 9\% more likely to succeed than those receiving no feedback. Test case feedback, by contrast, shows a slightly lower hazard (HR $= 0.92$, $p = 0.05$), consistent with the longer submission sequences we observe descriptively. Together, our results suggest that natural language feedback improves both completion rates and convergence speed, while test case feedback offers no reliable advantage in either dimension.

\subsection{Quality of Test Case Feedback}

We categorize AI-generated test cases as \textbf{valid} (fails the student's submission but passes the reference solution), \textbf{false-positive} (fails both), or \textbf{no-effect} (exposes no discrepancy). We find that 66\% of generated test cases are valid, 24\% are false positives, and 10\% have no effect. Submissions receiving valid test cases exhibit the highest average pass rates; false-positive and no-effect cases are associated with substantially lower performance. Importantly, we find that valid test cases are \emph{not} associated with fewer subsequent submissions---higher-quality feedback improves correctness without shortening the debugging process. We believe this reflects the limited systematic debugging skills of introductory programming students, who struggle to efficiently exploit a counterexample even when it is correct.

\subsection{Quality of Natural Language Feedback}

To assess feedback quality, we evaluate 442 natural language feedback instances using GPT-4.1 as a judge, labeling each as \emph{helpful}, \emph{vague}, \emph{incorrect}, or \emph{harmful}. We find that 92.5\% of instances are labeled helpful (409/442), 5.4\% vague (24/442), 2.0\% incorrect (9/442), and none harmful. Vague feedback typically offered general debugging suggestions without grounding them in the student's specific code; incorrect feedback reflected rare misattributions of the error source. Crucially, no feedback instance explicitly revealed a solution or bypassed student reasoning. Our results suggest that when quality failures occur, they are far more likely to involve insufficient specificity than pedagogically harmful guidance---supporting the safe deployment of natural language feedback in classroom settings.

\section{Discussion}

We find that natural language feedback is associated with both higher completion rates and faster convergence to correct solutions. We argue that this pattern is consistent with a cognitive load explanation: novice learners often struggle to interpret raw test failures, and natural language hints that translate program behavior into accessible explanations lower the barrier to identifying the next fix. Importantly, the significant hazard ratio in our survival analysis indicates that this effect reflects faster convergence, not merely a ceiling-driven success rate difference.

Test case feedback presents a more nuanced picture. We find its aggregate effect on outcomes to be weak, but we observe that this masks meaningful variation in feedback quality. Valid test cases are associated with higher pass rates, yet they do not reduce subsequent submission counts---suggesting that introductory students lack the systematic debugging skills needed to efficiently exploit a counterexample even when it correctly identifies a flaw. We believe this points to a fundamental mismatch between the form of the feedback and the skill level of the recipient, and motivates pairing test case feedback with explanatory guidance.

More broadly, our results highlight that evaluations of AI-assisted feedback should focus on learning trajectories rather than aggregate success rates, explicitly measure feedback quality, and interpret iteration counts cautiously---effective feedback may improve correctness without reducing submission effort.

\section{Limitations}

We acknowledge several limitations of this work. First, although feedback conditions were randomized within labs, delivery was conditionally triggered by incorrect submissions, making all analyses associative rather than causal. Second, we do not include expert-authored human feedback as a baseline, so we cannot assess how AI-generated modalities compare to high-quality instructor feedback. Third, our outcome measures---test case pass rates, time to correctness, and iteration patterns---capture short-term dynamics rather than longer-term retention, conceptual understanding, or transfer to new problems.

\section{Future Work}

Given these limitations, several directions remain open. One promising direction is developing models that predict test case invalidity before delivery, enabling real-time filtering or repair. Another is applying sequential and hierarchical models to characterize debugging pathways and persistence patterns across feedback conditions. More broadly, we hope \textsc{ProgFeed} can serve as a foundation for benchmarking AI feedback systems not only on output quality, but on their downstream effects on student learning in authentic classroom settings.

\section{Conclusion}

We presented a large-scale classroom study and accompanying dataset, \textsc{ProgFeed}, examining how different forms of AI-generated feedback affect student performance in an introductory programming course. Across 6{,}693 submissions from 215 students, we find that natural language hints are significantly associated with higher completion rates and faster convergence to correct solutions, while test case feedback yields weak aggregate effects that depend critically on feedback validity. Notably, even valid test cases do not reduce subsequent submission counts, suggesting that introductory students lack the debugging skills to efficiently exploit counterexamples---pointing to a mismatch between feedback form and learner readiness. Our quality analyses further reveal that natural language feedback is overwhelmingly helpful (92.5\%), while test case validity is far from guaranteed (66\%), underscoring that evaluating feedback quality---not merely its presence---is essential. Together, these findings argue that the \emph{form} of AI-generated feedback is not a superficial design choice but a determinant of pedagogical impact. We release \textsc{ProgFeed} to support further research on learning dynamics, feedback quality, and the responsible deployment of AI in educational settings.

\bibliographystyle{unsrtnat}
\bibliography{custom}

\end{document}